\documentclass[preprint,preprintnumbers,showpacs,amsmath,amssymb]{revtex4}

\usepackage{graphicx}

\usepackage{dcolumn}

\usepackage{bm}

\begin{document}

\title{Coherent oscillations in a superconducting multi-level quantum system}

\author{ J. Claudon$^{1}$, F. Balestro$^{1,2}$, F. W. J. Hekking$^{3}$, and O. Buisson$^{1}$}

\affiliation{$^{1}$Centre de Recherches sur les Tr\`es Basses Temp\'eratures,
laboratoire associ\'e \`a l'Universit\'e Joseph
Fourier, C.N.R.S., BP 166, 38042 Grenoble-cedex 9, France\\
$^{2}$Department of Nanoscience, Delft University of Technology, Lorentzweg 1, 2628 CJ Delft, The Netherlands\\
$^{3}$Laboratoire de Physique et Mod\'elisation des Milieux Condens\'es, CNRS \&
Universit\'e Joseph Fourier, BP 166, 38042 Grenoble-cedex 9, France}

\begin{abstract}
We have observed coherent oscillations in a multi-level quantum
system, formed by a current-biased dc SQUID. These oscillations
have been induced by applying resonant microwave pulses of flux.
Quantum measurement is performed by a nanosecond flux pulse that
projects the final state onto one of two different voltage states
of the dc SQUID, which can be read out. The number of quantum
states involved in the coherent oscillations increases with
increasing microwave power. The dependence of the oscillation
frequency on microwave power deviates strongly from the linear
regime expected for a two-level system and can be very well
explained by a theoretical model taking into account the
anharmonicity of the multi-level system.
\end{abstract}


\maketitle

Up to now, in view of quantum information processing, experiments
in solid state devices have concentrated only on the
implementation of two-level quantum systems. A variety of quantum
circuits based on Josephson
junctions~\cite{Makhlin01,Vion02,Martinis02,Chiorescu03}
and quantum dots~\cite{Hayashi03} have been proposed and
realized. Rabi oscillations have been observed showing that the
two lowest levels can be manipulated coherently. However, the
two-level system is not the only one useful for quantum
computation. For instance, multi-level systems are of interest for
solving database search problems using quantum
algorithms~\cite{Lloyd99}, as was demonstrated with Rydberg
atoms~\cite{Ahn00}. Recent theoretical proposals discuss the use
of multilevel quantum systems in solid state devices for quantum
information processing~\cite{Leuenberger01}. In
superconducting devices such as a current-biased Josephson
junction~\cite{Voss81} or a rf SQUID~\cite{Friedman00},
many experiments were performed demonstrating the multi-level
quantum nature of these systems. However no experiment has probed 
coherent behaviour yet. In this
letter, we report, to our knowledge, the first observation of coherent oscillations
in a multi-level solid-state circuit, based on Josephson
junctions. The non-linearity of the Josephson junction plays a
crucial role in the quantum dynamics of the device.

\begin{figure}
\resizebox{0.6\textwidth}{!}{\includegraphics{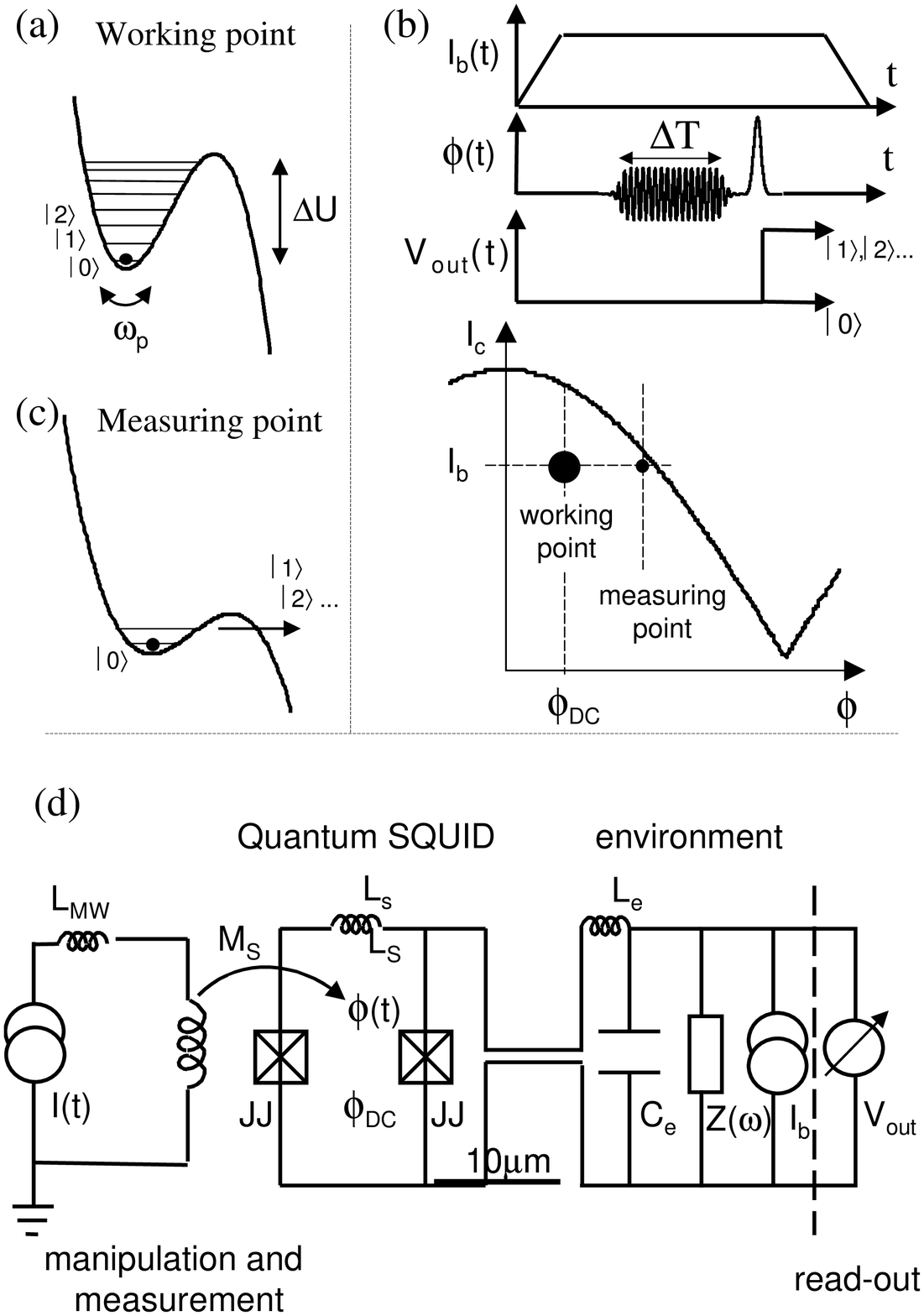}}
\caption{ Quantum experiment and measurement procedure in a dc
SQUID. (a) Illustration of the fictitious particle trapped in a
deep anharmonic well at the working point. (b) Experimental
procedure, as explained in the text. (c) The well at the measuring
point: during the nanosecond flux pulse, the particle
 can escape or remain in the well depending on its quantum state.
 (d) The dc SQUID consists of two identical Josephson
junctions coupled by an inductance $L_{s}$. The quantum circuit is
decoupled from the environment, symbolized by $Z({\omega})$, by a
large inductance $L_{e}$ and a capacitor $C_e$; it is coupled by a
mutual inductance $M_{s}$ to the MW pulse or nanosecond pulse
through a $50 \Omega$ coax line, terminated by an inductance
$L_{MW}$. \label{Proceduremesure}}
\end{figure}

Specifically, the quantum system that we study in our experiments
is a current-biased dc SQUID. It consists of two Josephson
junctions each with critical current $I_{0}$ and capacitance
$C_{0}$. The junctions are embedded in a superconducting loop of
inductance $L_{s}$, threaded by a flux $\Phi $. Since
$L_{s}I_{0}\approx \Phi_{0}/2\pi$, the phase dynamics of the two
junctions can be treated as that of a fictitious particle of mass
$m=2C_{\rm 0}(\Phi_{0}/2\pi)^2$ moving in a two-dimensional
potential, which contains valleys and
mountains~\cite{Seguin92,Li02,Balestro03}. Here $\Phi_{0} = h/2e$
is the superconducting flux quantum. The local minima are
separated from each other by saddle points, where the particle can
escape. Along the escape direction, the potential is cubic and can
be characterized by a frequency $\omega_{p}$ and a barrier height
$\Delta U$, see Fig.~\ref{Proceduremesure}a. These two quantities
depend on the magnetic flux and vanish at the critical current
$I_{c}$ of the SQUID. We assume a complete separation of the
variables along the escape direction and the transverse one by
neglecting the coupling terms between these two directions. In
this approximation, the dynamics of the SQUID's phase $\phi$ along
the escape direction is similar to the dynamics of the phase of a
current-biased single Josephson
junction~\cite{Martinis87,Larkin86}. The parameters $I_{c}$,
$\Delta U$ and $\omega_{p}$ are renormalized, thereby taking into
account the two-dimensional nature of the potential, as it was
demonstrated in~\cite{Seguin92,Balestro03}. For bias currents
$I_{b} < I_{c}$, the particle is trapped in a local minimum and
its quantum dynamics is described by
\begin{equation}
\hat{H}_{0}=\frac{1}{2}\hbar\omega _{p} (\hat{P}^2+\hat{X}^2)- \sigma\hbar\omega _{\rm
p}\hat{X}^3 , \label{Hamiltonian}
\end{equation}
where  $\hat{P}=(1/\sqrt{m \hbar \omega _{\rm p}}) P$ and
$\hat{X}=(\sqrt{m \omega _{\rm p}/\hbar}) \phi$ are the reduced
momentum and position operators, respectively. Here, $P$ is the
operator conjugate to $\phi$;  $\sigma =1/(6a) [2(1-I_{\rm
b}/I_{\rm c})]^{-5/8}$  is the relative magnitude of the cubic
term compared to the quadratic (harmonic) term. The parameter $a$
is a constant, $a\sim 11$ for our SQUID. For $I_{b}$ well below
$I_{c}$, $\Delta U \gg \hbar \omega_{p}$ and many low-lying
quantum states are found near the local minimum. These states,
describing the oscillatory motion within the anharmonic (cubic)
potential, are denoted $|n\rangle$ for the $n$th level, with
$n=0,1,2, \ldots$. The corresponding energies $E_{n}$ were
calculated in Ref.~\cite{Larkin86}. Tunnelling through the barrier
can be neglected for the lowest-lying states with $n\hbar\omega
_{\rm p} \ll \Delta U$. The effect of a time-dependent flux
$\Phi(t)$ can be included in Eq.~(\ref{Hamiltonian}) by adding the
time-dependent term $\alpha (t) \hbar\omega _{\rm p}\hat{X}$ where
$\alpha (t)$is proportional to $\Phi(t)$~\cite{Buisson03}. In our
experiment, deep quantum states are excited by applying microwave
(MW) flux pulses characterized by their frequency $\nu$, amplitude
$\Phi_{MW}$ and duration $\Delta T$. Starting with $|\Psi (0)
\rangle=|0\rangle$, the state evolves with the MW pulse of
duration $\Delta T$ into a coherent superposition $|\Psi(\Delta T)
\rangle =a_{0}|0\rangle+a_{1}|1\rangle+\ldots+a_{n}|n\rangle +
\ldots$

Our procedure to perform quantum experiments consists of four
successive steps as depicted in Fig.~\ref{Proceduremesure}b. A
bias current $I_{b}$ is switched on through the SQUID at fixed
magnetic flux $\Phi_{dc}$ to prepare the circuit at the working
point in the initial state $|0\rangle$. The application of a MW
flux pulse produces the superposition $|\Psi(\Delta T) \rangle$.
Then, a flux pulse of nanosecond duration is applied which brings
the system to the measuring point such that the state
$|\Psi(\Delta T) \rangle$ is projected onto either the zero or the
finite voltage state of the SQUID (Fig.~\ref{Proceduremesure}c).
The result of this quantum measurement can be read out by
monitoring the voltage $V_\mathrm{out}$ across the dc SQUID.
Finally, $I_{b}$ is switched off such that the circuit is reset
and the experiment can be repeated to obtain state occupancy.

The aforementioned quantum measurement procedure was discussed
theoretically by us in Ref.~\cite{Buisson03}. Differently from
previous experiments which used current bias
pulses~\cite{Vion02,Chiorescu03} or MW pulses~\cite{Martinis02},
we implemented a quantum measurement using a large-amplitude flux
pulse with nanosecond duration. This flux pulse reduces the SQUID
critical current to a value very close to the bias current such
that $\Delta U\sim\hbar\omega_{p}$. Ideally, the pulse with
optimal amplitude projects the excited states with $n>0$ onto the
voltage state ($V_\mathrm{out}$ is twice the superconducting gap);
the ground state $|0\rangle$ is projected onto the zero-voltage
state. As the SQUID is hysteretic, the zero and finite voltage
states are stable. The efficiency of this one-shot measurement is
estimated to be 96\%~\cite{Buisson03}. In the experiment, we use a
pulse of $2.5 ns$ duration whose $1.5 ns$ rise and fall times are
long enough to guarantee adiabaticity: transitions between
$|0\rangle$ and excited states are estimated to occur with a
probability less than 1\% for a typical pulse amplitude of
0.06$\Phi_{0}$. The measurement pulse can be applied with a
variable delay after the end of the MW pulse. In the measurement
procedure, this delay is kept as short as $1.5 ns$ to limit
relaxation processes. The result of the quantum measurement is
obtained by measuring the escape probability $P_{esc}$ repeating
the experiment up to about 4000 times.

The measured dc SQUID consists of two aluminum Josephson junctions
with $I_{0}=3.028 \mu A$ and $C_{0}=0.76pF$, coupled by an
inductance $L_{s}=98pH$ (see Fig.~\ref{Proceduremesure}d). The
quantum circuit is decoupled from the external classical circuit
by long on-chip superconducting thin wires of large total
inductance, $L_e=15nH$, and a parallel capacitor, $C_e=150pF$. The
quantum circuit and the long superconducting wires are realized by
e-beam lithography and shadow evaporation. The
nominal room temperature microwave signal is guided by $50 \Omega$
coax lines and attenuated twice by 20dB (at 1.5K and 30mK,
respectively) before reaching the SQUID through a mutual
inductance $M_{s}=0.7pH$. The MW line is terminated by an
inductance estimated to be $L_{MW}=1nH$. Special care was taken to
avoid spurious environmental microwave resonances. The chip is
mounted in a shielded copper cavity, cooled down to about 30 mK,
whose cut-off frequency is above 20 GHz. Moreover, the thin film
capacitor, $C_e$, close to the chip decouples the microwave signal
from the low-frequency bias lines.

The escape probability is first obtained by measurements using
long (duration $\Delta t= 50 \mu s$) pulses of the bias current
$I_{b}$. From the dependence of the escape current (defined as the
current at which $P_{esc}=1/2$) on $\Phi_{dc}$, the experimental
parameters of the SQUID are extracted~\cite{Balestro03}. At the
maximum value of $I_{c}$ obtained for $\Phi_{dc}/\Phi_{0}=-0.085$,
the measured escape rate can be fit by the well-known MQT
formula~\cite{Li02,Balestro03} yielding the same SQUID parameters
within a 2 \% error. This result confirms that the circuit remains
in the ground state when only $I_{b}$ pulses are applied. For
other values of $\Phi_{dc}$, the width of the escape probability
is larger than the MQT prediction by a factor up to 4 indicating a
residual low-frequency flux noise in our
sample~\cite{Chiorescu03}. This noise reduces the efficiency of
our quantum measurement since it smears the escape probability
difference  between the ground state and the excited states. In
the experiment the nanopulse amplitude is adjusted in order that
the escape from $|0\rangle$ is close to 1\%. For such a pulse
amplitude at $\Phi_{dc}/\Phi_{0}=0.095$, we have estimated that
the escape probabilities of the excited states $|1\rangle$,
$|2\rangle$, $|3\rangle$ and higher states are about 30 \%, 60 \%,
90 \% and 100\% respectively. The sensitivity to flux noise is
weakest at the "optimal" point $\Phi_{dc}/\Phi_{0}=-0.085$;
however, the SQUID's sensitivity to MW and measurement pulses is
also weak such that this point cannot be used for accurate
measurements.

\begin{figure}
\resizebox{0.8\textwidth}{!}{\includegraphics{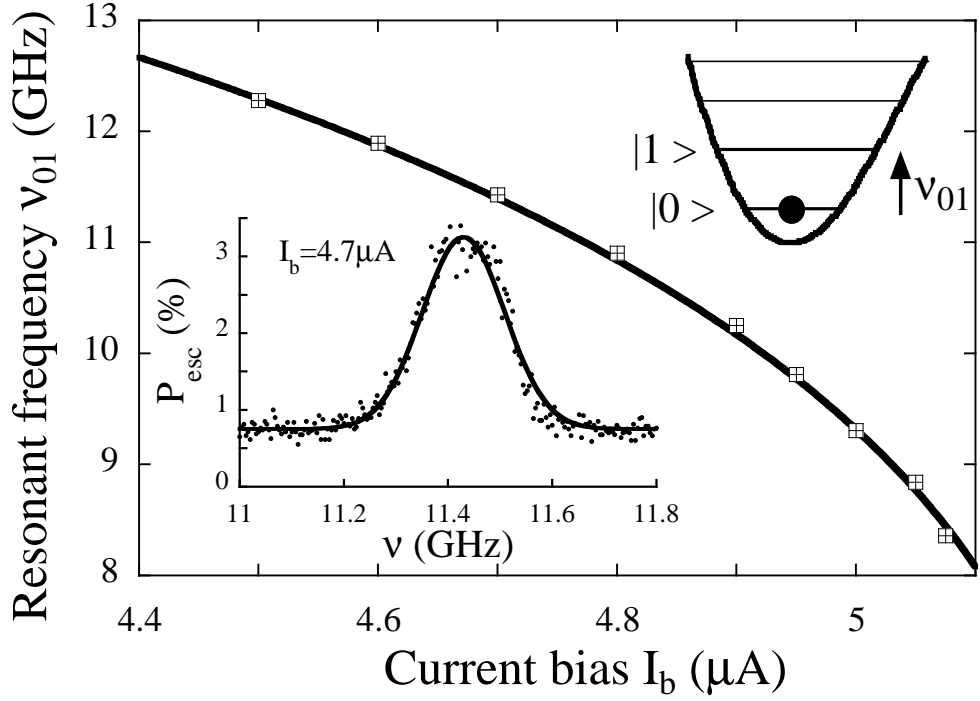}}
\caption{Resonant transition frequency $\nu_{01}$ between states
$|0\rangle$ and $|1\rangle$ as function of the bias current at
$\Phi_{dc}/\Phi_{0}=0.095$. The experimental dependence of
$\nu_{01}$ (symbols) on $I_{b}$ is perfectly described by the
semiclassical model \cite{Larkin86} (solid line). Spectroscopy is
performed by measuring the escape probability $P_{esc}$ versus
applied frequency $\nu$ (points in inset) at microwave power
A=-48dBm where A is the nominal microwave power at room
temperature. The fit to a Gaussian (solid line) defines the
resonant frequency $\nu_{01}=11.43 GHz$ and the width
$\Delta\nu_{01}=180 MHz$. \label{Resonance}}
\end{figure}

Spectroscopic  measurements were performed by sweeping the frequency of a MW flux
pulse of 25$ns$ duration in the low power limit. In the inset of Fig.~\ref{Resonance}
we show the corresponding resonance peak found for $P_{esc}$ versus microwave
frequency $\nu $. The resonant frequency $\nu
_{01}$ depends on $I_{b}$ as shown in Fig.~\ref{Resonance}; this dependence can be
very well fit by the semiclassical formula for the cubic potential~\cite{Larkin86}.
The parameters extracted from this fit are consistent within 2\% error with the
parameters extracted from the critical current versus flux dependence or from the MQT
measurements at $\Phi_{dc}/\Phi_{0}=-0.085$. The large width $\Delta\nu_{01}=180 MHz$
of the resonance peak is consistent with the presence of a residual low-frequency flux
noise. Indeed, since the flux through the SQUID fluctuates slowly, the expected
resonant frequency changes from one measurement to the other.

We also measured the lifetime of the first excited states,
analyzing the decay of the resonance height. Upon increasing the
delay time between the end of the MW pulse and the nanosecond dc
measurement pulse, the peak height is found to decay with times
ranging from 14ns to 60ns. The frequency broadening associated
with these lifetimes is estimated to be smaller than 10MHz, i.e.,
not only less than the width $\Delta\nu_{01}$ of the resonance
peak but also less than the detuning frequency associated with the
anharmonicity between adjacent levels.

\begin{figure}
\resizebox{0.8\textwidth}{!}{\includegraphics{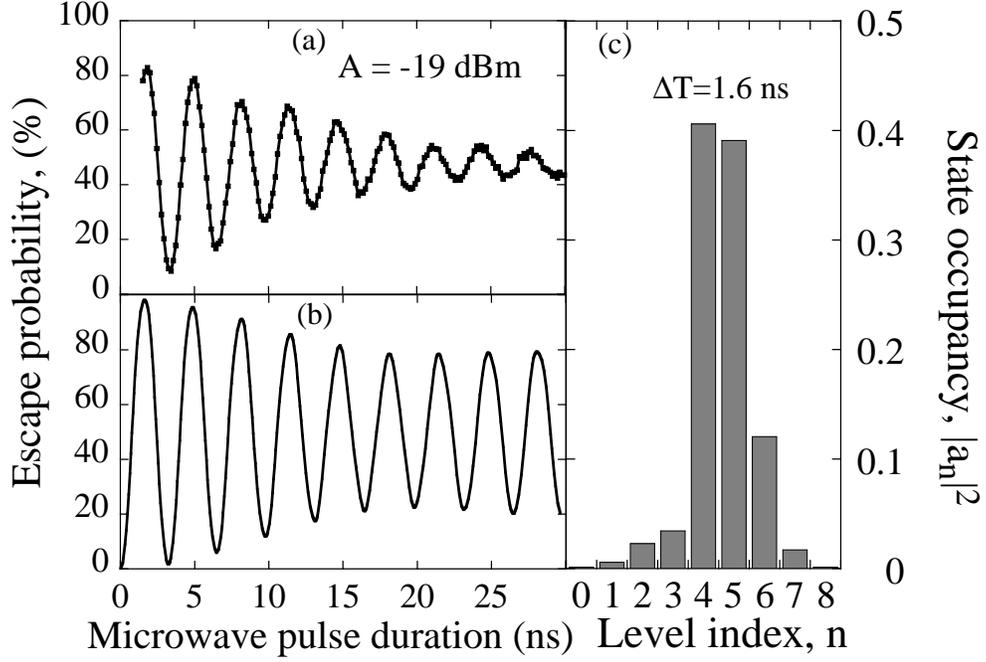}}
\caption{a) Measured escape probability versus MW pulse duration at the working point, defined by
$\Phi_{dc}/\Phi_{0}=0.095$ and $I_{b}=4.7\mu A$, with about 15
levels localized in the anharmonic well. b) Predicted escape
probability for a MW amplitude $\Phi_{MW}/\Phi_{0}=0.002$. c)
$a_{n}$ coefficients of the state obtained after $1.6ns$ duration
of the MW pulse corresponding to the first maximum of the coherent
oscillations. \label{Coherentoscillation}}
\end{figure}

\begin{figure}
\resizebox{0.8\textwidth}{!}{\includegraphics{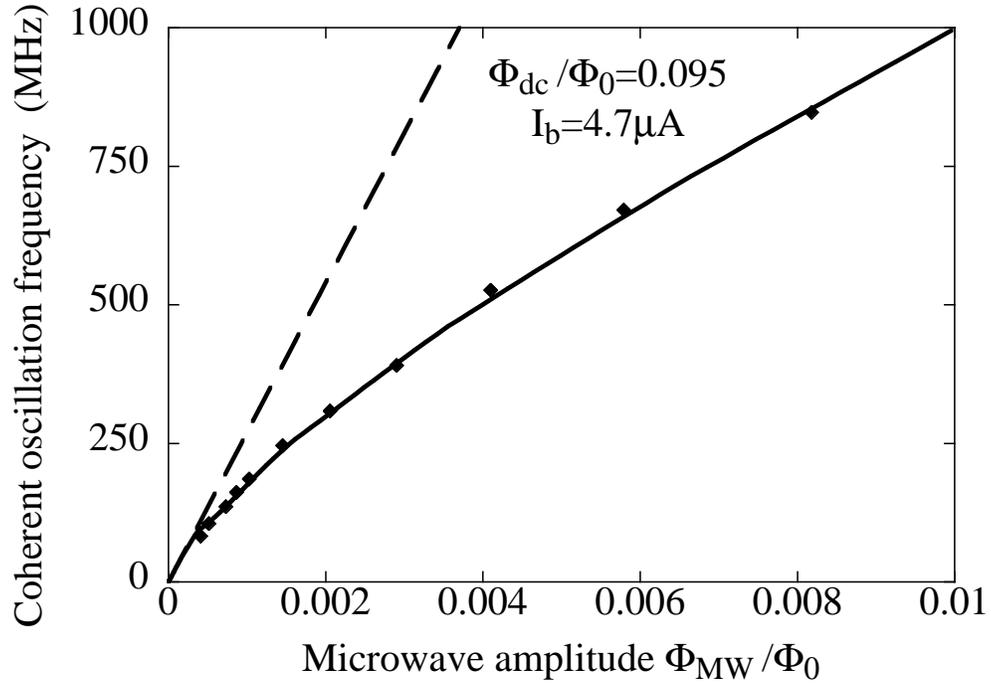}}
\caption{Frequency of coherent oscillations versus the normalized
MW flux amplitude. Dots are experimental results, the solid line
the theoretical prediction for a multi-level anharmonic quantum
system. Dashed line is the Rabi theory for a two-level system.
Experimental calibration of $\Phi_{MW}$ is deduced from the fit
and is consistent with the estimated $\Phi_{MW}$ applied to the
set-up. \label{RabiversusVrf}}
\end{figure}

By applying short MW pulses at the resonant frequency $\nu = \nu
_{01}$, we induce coherent quantum dynamics. Rabi-like coherent
oscillations were observed by measuring escape probability versus
MW pulse duration $\Delta T$. In Fig.~\ref{Coherentoscillation}a
the escape probability oscillates at a frequency $\Omega_{coh}/2\pi$ of
about 300MHz. This frequency increases as the MW flux power
increases and ranges from about 100 MHz to 1000 MHz in our
experiment. It is always much smaller than the resonance frequency
$\nu _{01}$. Oscillation amplitudes as large as 70\% were observed
for the largest MW power. The oscillations are damped with a
characteristic attenuation time of about $12ns$. Similar coherent
oscillations have been observed at different working points.  The
dependence of $\Omega_{coh}$ as a function of the MW amplitude
$\Phi_{MW}$ is shown in Fig.~\ref{RabiversusVrf}. The linear
dependence predicted by Rabi theory for a two level
system~\cite{Rabi37} is not observed in our measurements.

To analyse the observed coherent oscillations, we use the
following model. We ignore relaxation and decoherence processes
and consider the applied MW frequency to be the resonant frequency
$\nu _{01}$. If we furthermore assume the MW pulse to couple the
eigenstate $|n\rangle$ to its nearest-neighbor levels only, the
time-dependent part of the Hamiltonian reads $ \alpha (t)
\hbar\omega _{\rm p}\hat{X} \approx
 b\frac{\Phi_{MW}}{\Phi_{0}} \cos( 2 \pi \nu_{01} t)\hbar \omega _p \sum _n
\sqrt{n/2} [|n\rangle\langle n-1| + |n-1\rangle\langle n| ], $
where $b \sim 34$ at the considered working point. As $\Omega_{coh}/2\pi\ll \nu_{01}$, we can use the
rotating wave approximation and treat this term non-perturbatively.
We obtain the coefficients $a_{n}$ of the coherent superposition
$|\Psi(\Delta T) \rangle$ generated by a pulse of duration $\Delta
T$: $a_{n}=\sum _k \langle e_{k}|0\rangle\langle n|e_{k}\rangle
e^{-i\lambda_{k}\Delta T/\hbar}$, where $|e_{k}\rangle$ and
$\lambda_{k}$ are the eigenstates and eigenvalues of the
Hamiltonian in the rotating frame. The latter are determined
numerically and then used to calculate the probabilities
$|a_n(\Delta T)|^2$, whose oscillatory time dependence is
determined by the frequency differences $\lambda _k - \lambda _j$.
At very low MW amplitude ($ b\Phi_{MW}/\Phi_{0}\hbar \omega_{p}\ll
2E_{1}-(E_{2}+E_{0})$), we obtain the two-level limit:
$|\Psi(\Delta T) \rangle$ oscillates between $|0\rangle$ and
$|1\rangle$ at the Rabi frequency
$\Omega_{coh}=b\omega_{p}/\sqrt{2}\Phi_{MW}/\Phi_{0}$~\cite{Rabi37}.
At larger MW amplitude, an increasing number of
states is predicted to participate in the oscillations~\cite{Amin04}. As an
example, in Fig.~\ref{Coherentoscillation}b the theoretical escape
probability is plotted as function of MW pulse duration taking
into account the finite efficiency of our measurement. The MW
amplitude $\Phi_{MW}/\Phi_{0}=0.002$ was adjusted to fit the
measured oscillation frequency. The MW power calibration found
from the fit is consistent, within a 30 \% error, with the actual
applied MW amplitude and with the microwave line and coupling
parameters, which were measured independently. The model
calculation predicts very large amplitudes, consistent with the
experimental results. At $\Delta T= 1.6ns$ the amplitude is close
to 100\%. In Fig.~\ref{Coherentoscillation}c we indicate the
occupancies $|a_n|^2$ of the states $|n\rangle$
that participate in the coherent superposition state at this time. We see that
it is concentrated mainly on the states $|4\rangle$, $|5\rangle$
and $|6\rangle$. Finally, the model predicts a slight beating in
the oscillations at $\Delta T\sim 20ns$ where the amplitude
reaches a minimum. These beating phenomena will not be analyzed in
our experiment because they occur at time scales of the order of
the above-mentioned attenuation time. For short
duration times, the
predicted coherent oscillation frequency is given by
$\mbox{min}_{k,l}\{(\lambda_{k}-\lambda_{l})/\hbar \}$. In this limit and 
using the
MW calibration deduced from Fig.~\ref{Coherentoscillation}, we calculate $\Omega_{coh}/2\pi$ as a function
of $\Phi_{MW}/\Phi_{0}$. The result is plotted in Fig.\ref{RabiversusVrf} together with the experimental
one. The perfect agreement between them shows that our model captures the physics of the coherent oscillations.

We reported, to the best of our knowledge, the first observation
of coherent oscillations in a multi-level solid-state-based
integrated circuit. In our experiment, we have induced
coherent superpositions of quantum states which involves many levels using monochromatic
microwaves. In order to perform the quantum measurement, we
implemented a new measurement procedure based on nanosecond flux
pulse through the SQUID. The agreement between our results
and the theory proves that the essential physics of our non-linear
quantum circuit is well understood.

We thank B. Camarota, F. Faure, Ph. Lafarge, L. L\'evy, D. Loss,
M. Nunez-Regueiro, B. Pannetier, J. Pekola, and A. Ratchov for
useful discussions. This work was supported by ACI and ATIP
programs, and by the Institut de Physique de la Mati\`ere
Condens\'ee. FH acknowledges support from Institut Universitaire
de France and the hospitality of NTT Basic Research Laboratories.

\newpage

\newpage

\newpage

\newpage

\end{document}